\documentclass[ %
12pt,
a4paper,
ngerman,
english,
]{scrartcl} %

\usepackage{babel}              
\usepackage[utf8]{inputenc}   
\usepackage[T1]{fontenc}        
\usepackage[scaled]{helvet}     
\setcapindent{0em}

\usepackage{amsmath}
\usepackage{amsfonts}
\usepackage{amssymb}
\usepackage{amsthm}    
\usepackage{graphicx}   
\usepackage{caption}
\usepackage{subcaption}
\usepackage{geometry} 
\usepackage[round,comma,sort]{natbib} 
\usepackage{color}
\usepackage{epstopdf}           
\usepackage{fancybox}           
\usepackage{tabu}
\usepackage{hhline}
\usepackage[draft]{hyperref}
\hypersetup{final}
\usepackage{cleveref}
\usepackage{makeidx}            
\usepackage{enumerate}          
\usepackage{float}
\usepackage{doi}
\makeindex

  \usepackage{pgfplots}
  \pgfplotsset{compat=newest}
  \usetikzlibrary{plotmarks}
  \usetikzlibrary{arrows.meta}
  \usepgfplotslibrary{patchplots}
  \usepackage{grffile}

\newif\ifPics %
\Picstrue 

\begin{document}
\title{Sensitivity and depth of investigation from Monte Carlo ensemble statistics} 

%
\newcommand*\samethanks[1][\value{footnote}]{\footnotemark[#1]}

\author{Christin Bobe%
  \thanks{ %
    Corresponding Author: christin.bobe@uni-osnabrueck.de, %
    Department of Environment, %
    Ghent University, %
    Gent, %
    Belgium, %
    Now at: Universit{\"a}t Osnabr{\"u}ck, %
    Neueste Geschichte und Historische
Migrationsforschung, %
	Osnabr{\"u}ck, %
	Germany
  } %
  \and Johannes Keller%
  \thanks{ %
    Institute for Applied Geophysics and Geothermal Energy, %
    RWTH Aachen University, %
    Aachen, %
    Germany, %
    Now at: Forschungzentrum J{\"u}lich, %
    Institute of Bio- and Geosciences, %
    IBG-3 (Agrosphere), %
    J{\"u}lich, %
    Germany %
    and 
    Centre for High-Performance Scientific Computing in
    Terrestrial Systems (HPSC-TerrSys), %
    Geoverbund ABC/J, %
    J{\"u}lich, %
    Germany
  } %
  \and Ellen Van De Vijver%
    \thanks{ %
    Department of Environment, %
    Ghent University, %
    Gent, %
    Belgium
  } %
} %

\date{} %
\maketitle

\begin{center}
\textbf{Accepted for publication in \textit{Geophysical Prospecting}. Further reproduction or electronic distribution is not permitted.}
\end{center}




\newpage

\begin{abstract}
For many geophysical measurements, such as direct current or electromagnetic 
induction methods, information fades away with depth. %
This has to be taken into account when interpreting models estimated from such measurements. %
For that reason, a measurement sensitivity analysis and determining 
the depth of investigation are standard steps during geophysical 
data processing. %
In deterministic gradient-based inversion, the most used sensitivity measure, the differential sensitivity, is readily available since these inversions require the computation of Jacobian matrices. %
In contrast, differential sensitivity may not be readily available 
in Monte Carlo inversion methods, since these methods do not 
necessarily include a linearization of the forward problem. %
Instead, a prior ensemble is used to simulate an ensemble of forward 
responses. %
Then, the prior ensemble is updated according to Bayesian inference. %
We propose to use the covariance between the prior ensemble and the 
forward response ensemble for constructing sensitivity measures. %
In Monte Carlo approaches, the estimation of this covariance does 
not require additional computations of the forward model. %
Normalizing this covariance by the variance of the prior ensemble, 
one obtains a simplified regression coefficient. %
We investigate differences between this simplified regression 
coefficient and differential sensitivity using simple forward 
models. %
For linear forward models, the simplified regression 
coefficient is equal to differential sensitivity, except for the 
influences of the sampling error and of the correlation structure 
of the prior distribution. %
In the non-linear case, the behaviour of the simplified regression coefficient as sensitivity measure is analysed for a simple non-linear 
forward model and a frequency-domain electromagnetic forward model. %
Differential sensitivity and the simplified regression coefficient are 
similar for prior intervals on which the forward model response is 
approximately linear. %
Differences between the two sensitivity measures increase with the 
degree of non-linearity in the prior range. %
Additionally, we investigate the correlation between prior ensemble 
and forward response ensemble as sensitivity measure. %
Correlation yields a normalised version of the 
simplified regression coefficient. %
We propose to use this correlation and the simplified regression 
coefficient for determining depth of investigation in Monte 
Carlo inversions. %
\end{abstract}

\section{Introduction}
\label{sec:introduction}


Many measurements used in applied geophysics, such as measurements 
from electrical resistivity tomography and electromagnetic induction methods, can be reproduced by assuming diffusive energy propagation. %
In combinations with simulations, such measurements allow to estimate 
subsurface distributions of physical properties, such as electrical 
conductivity, using inverse modeling techniques. %
There are two major types of inversion techniques, deterministic 
gradient-based and stochastic Monte Carlo (MC) inversions 
\cite[]{aster2018parameter}. %
Both inversion types rely on the influence of proposed parametric subsurface 
models on simulation responses \cite[]{mcgillivray1990methods}. %
The so-called measurement sensitivity characterises this influence 
by linearised analysis, i.e. partial derivatives of the simulated 
data with respect to model parameters. %
Using sensitivity, the influence of the different measurement 
signals can be compared. %
In particular, model parameters to which the measurements have negligible sensitivity can be identified. %

For a typical surface measurement, sensitivity is negligible from 
a characteristic depth downwards. %
This characteristic depth is called the depth of investigation (DOI) of this particular surface measurement. %
An estimation of the DOI is crucial as it can prevent over- or 
misinterpretation of the inversion results 
\cite[]{oldenburg1999estimating}. %
A determination of the DOI is not trivial if data from diffusive 
methods are processed, since (1) a depth of absolute zero sensitivity 
does not exist, and (2) sensitivity computations are usually 
restricted to parameter variations around a (deterministic) inversion 
result. %
If the inversion result is significantly different 
from the true subsurface parameters, the estimation of the DOI may 
be wrong. %

Due to the lack of a depth of truly zero sensitivity, a definition of the 
DOI always contains some degree of arbitrariness. %
Often, a rather small sensitivity threshold is defined relative to 
a reference sensitivity, for example using 5 $\%$ of the 
maximum sensitivity. %
Other approaches to DOI estimation define global sensitivity 
thresholds. %
For example, \cite{vest2012global} introduced the method of cumulative 
sensitivity. %
They compute all cumulative sums of sensitivities throughout a 
one-dimensional model starting from the bottom. %
This way, it is possible to give a global sensitivity threshold for 
DOI estimation, but the arbitrariness of this threshold remains. %
As the cumulative sensitivity is based on differential sensitivities, 
it must be assumed that the derived inverse image is a good 
representation of subsurface reality. %

Despite the helpful guidelines for DOI estimation pointed out in the 
previous paragraph, the problem associated 
with investigating the sensitivity only around the final inverse 
model remains for deterministic inversion results. %
This non-uniqueness problem was already considered in the early days 
of geophysical inverse processing, for example using the Backus-Gilbert 
analysis (e.g., \cite{backus1968resolving} and 
\cite{backus1967numerical}). %
However, in their analysis, only a linear range around the result of the 
inverse method is searched for subsurface models that equally well satisfy 
the observed data. %
As a first step towards coping with this non-uniqueness problem in 
fully non-linear deterministic inversions, 
\cite{oldenburg1999estimating} introduced the so-called DOI index 
for the interpretation of direct current and induced polarisation 
inversion results. %
This procedure was later adapted for the evaluation of electromagnetic 
inversion results (e.g., \cite{brosten2011inversion}). %
To derive the DOI index, the reference model for 
the inversion is altered to expose features in the inverse image that 
strongly depend on the choice of the reference model. %
However, alteration is often limited to two different reference models, 
leaving large parts of the model parameter space unexplored. %


As an alternative to deterministic methods, stochastic Monte Carlo (MC) inversion 
methods have been used used more frequently in recent years as 
the available computational resources are growing 
\cite[]{tarantola2005inverse}. %
Such methods solve the parameter estimation problem by random sampling of probability distributions and searching for an inverse solution given as a random parameter vector that follows a probability distribution as first formalized by \cite{tarantola1982inverse}.
MC methods tackle the general non-uniqueness problem by performing 
an as extensive as possible search of the model parameter space using 
random sampling. %
In this way, MC methods allow to characterise the uncertainty of the inversion result. %
Several variations of MC approaches exist and are used in geophysical 
practice.  %
A thorough compilation, featuring popular applications of Markov chain Monte Carlo methods \cite{mosegaard1995monte}, can be found in \cite{sambridge2002monte}. %
Although, since the publication of their review, several new MC methods have been introduced, for example trans-dimensional, multi-chain or approximate Bayesian methods (\cite{malinverno2002parsimonious}, \cite{sambridge2006trans}, \cite{socco2008improved}, \cite{vrugt2009accelerating}, and \cite{Bobe2019KEG}). %

In MC methods, the model parameter space that is sampled is defined based on 
prior knowledge. %
This prior knowledge can be understood in the Bayesian sense. %
It also serves as implicit regularization in such probabilistic frameworks. %
With its subjectivity, the definition of the prior is often the 
main criticism of Bayesian inversion methods \cite[]{scales2001prior}. %
However, in the following we will assume that defined prior distributions do reflect 
possible realizations of the subsurface according to prior knowledge. %

Applying MC sampling, one has to compute the forward response for each prior 
model realization. %
Drawing a large number of prior samples, one obtains detailed insight on 
how changes in the parameter model relate to changes in the measurement 
response. %
Above, we defined such a relation as measurement sensitivity. %
Deriving covariance of the prior and the forward response ensemble and dividing 
this by the prior ensemble variance, we derive a simplified 
regression coefficient (SimRC). %
For dimensionless comparison of signals, the SimRC can be normalised by 
the variance of the forward response ensemble, yielding the 
correlation of prior and forward response ensemble. %

Inversion updates derived from MC analysis are conditioned on the above 
mentioned covariances and the actual measurement data. %
Therefore, MC analysis does not require a linearised analysis of the forward 
problem and consequently, the traditionally used differential 
sensitivities are not readily available. %
Thus, commonly used DOI methods (see above) cannot immediately be 
applied. %
Moreover, local differential sensitivities have only limited informative value 
when evaluating results from global MC analysis and are usually not 
computed in such (e.g., \cite{minsley2011trans}). %

For DOI estimations in MC inversions, the inverted model is often cut at a 
depth at which posterior uncertainty becomes 'too large' to be considered 
informative (e.g., \cite{brodie2012transdimensional}). 
One inconvenience of this DOI selection procedure is that the absolute 
size of the posterior uncertainty interval for model 
sections to which the measurements have negligible sensitivity almost only depends on the chosen prior probability. %
When a prior probability incorporates a high degree of certainty, 
a high degree of certainty will as well be seen for the posterior model, also for 
model parameters to which the measurements have vanishing sensitivity. %
This behaviour of the posterior probability makes a DOI threshold based 
on the size of the posterior uncertainty inevitably conditioned on 
the chosen prior distribution. %
Here, it should be noted that model sections with vanishing sensitivity 
or correlation to the measurements can certainly be considered just as 
reliable as the prior knowledge. %
Nevertheless, practitioners usually have keen interest in exploring the 
sensitivity of their model and in delineating their prior 
knowledge from the information gained from the inversion update. %

One method for such a delineation is the computation of the Kullback-Leibler (KL) divergence. %
The KL divergence was recently presented for Bayesian inversion results in \cite{blatter2018trans}. %
The KL divergence gives the difference in information content of prior and posterior distribution. %
At a depth below prior and posterior distribution are near equal, no information was added by the measurement data. %
This depth can be interpreted as a DOI. %
However, the KL divergence is a purely statistical measure, allowing no analysis of measurement sensitivity and requires that a posterior distribution is available. %

Analysis of the sensitivity not only gives 
insight in the nature of the inversion 
update, it can also support reconsideration of measurement design. %
However, instead of additionally computing differential sensitivities, 
we propose the use of the SimRC and of the correlation for model 
sensitivity analysis. %
The closely related regression coefficient as used by 
\cite{saltelli2004sensitivity} is often considered as a global 
sensitivity and is informative for the sampled model parameter space 
\cite[]{saltelli2004sensitivity}. %
Additionally, the correlation allows for comparison of different measurement 
signals, and a vanishing correlation gives an intuitive measure for the 
DOI. %


The text at hand treats the estimation of sensitivity and DOI for 
probabilistic MC inversion methods using the SimRC and correlation 
analysis. %
For some non-probabilistic MC methods (optimization methods), the presented 
work might be applicable, however, these will not be considered here. %
We will describe (1) how the readily available 
ensembles of prior distribution and forward response can be used to estimate 
SimRC sensitivity, and (2) how the SimRC and the corresponding 
correlation functions can be used to estimate a DOI. %
This DOI distinguishes between updated model parameters and parameters 
that essentially remain described by prior information. %
Additionally, common DOI estimation and illustration methods (e.g., \cite{vest2012global} and 
\cite{oldenburg1999estimating}) can be applied to the SimRC functions. %

The manuscript is structured as follows. %
First, we show how the SimRC is derived for Monte Carlo style 
inversions.
Second, we explain how the SimRC relates to differential sensitivities in 
linear inverse problems. %
We also show how the SimRC and derived correlation functions 
can be used when estimating the DOI in surface measurement data 
inversions. %
Further, we include three synthetic studies illustrating differences and similarities between the differential sensitivity and the SimRC. %
We introduce a linear and a non-linear toy model to provide a 
semi-quantitative analysis of the influence of the prior distribution 
(sampling) specifications on the SimRC, namely prior correlation, 
non-linearity of the forward model, and the sample size. %
For the application to a geophysical problem, we provide simulations 
of a one-dimensional frequency-domain electromagnetic forward model in which 
different measurement signals with different differential 
sensitivities to the subsurface model are compared and analysed using 
the SimRC sensitivity. %

\section{Theory}
\label{sec:theo}

In the following paragraphs we recall the general concept of Bayesian 
inference in Monte Carlo (MC) inversion approaches. %
Hereby, we will focus on the computation of the sample covariance 
between the prior and forward response ensemble. %
We motivate constructing measurement sensitivity measures from 
this covariance. %
The first proposed sensitivity measure is the covariance normalised 
by prior variance, a simplified regression coefficient (SimRC). %
We show that the SimRC and differential sensitivity are equal for the case 
of a linear forward model and no correlation in the prior distribution. %
Additionally, we propose using the SimRC and the correlation 
between prior ensemble and forward response ensemble for estimating the 
depth of investigation (DOI) of a geophysical surface measurement. %

\subsection{Bayesian inference}

In Bayesian inversion approaches, prior information is represented in 
a probabilistic manner. %
According to Bayes' theorem, this prior information is combined with 
the observed data to derive a posterior probability density function (PDF). %
The prior distribution is expressed as the PDF 
$\rho(\textbf{m})$ for the random vector of model parameters 
$\textbf{m} \in \mathbb{R}^{n_{par}}$, where $n_{par}$ is the number 
of model parameters. %
The likelihood function is the PDF $\rho(\textbf{d}|\textbf{m})$ 
that denotes the conditional probability that a set of observation 
$\mathbf{d} \in \mathbb{R}^{n_{obs}}$ is measured for a given set of parameters. %
Here, $n_{obs}$ denotes the number of observation variables. %
For geophysical measurements, the likelihood gives information on whether 
an observation is compatible with predicted responses $g(\textbf{m})$, 
where $g$ is a geophysical forward model. %
The posterior PDF for the random vector of model parameters is derived 
by applying Bayes' theorem:

\begin{equation}
\rho(\textbf{m}|\textbf{d}) \propto \rho(\textbf{d}|\textbf{m}) \rho(\textbf{m}).
\label{eq:Bayestheorem}
\end{equation}

For large-dimensional non-linear inverse problems, no analytical solution 
to Equation \ref{eq:Bayestheorem} can be given 
\cite[]{tarantola2005inverse}. %
Numerical sampling of the involved probability distributions is 
one possibility for obtaining an approximate solution to Bayes' theorem. %
Monte Carlo methods are a class of algorithms based on numerical 
random sampling. %
Collecting the Monte Carlo samples in matrices, we approximate 
$\rho(\textbf{m})$ by a matrix $\mathbf{M} \in \mathbb{R}^{(n_{par} \times N)}$
\cite[]{evensen2003ensemble}, where $N$ is the number of samples. %
For each column in $\mathbf{M}$, forward responses are simulated. %
The resulting forward response ensemble is collected in a 
matrix $\mathbf{G}=g(\mathbf{M})\in \mathbb{R}^{(n_{obs} \times N)}$. %

In principle, drawing a large number of samples $N$ the model 
parameter space is searched nearly exhaustively covering almost all 
possible parameter realizations. %
In comparison with deterministic inversion, non-uniqueness problems 
are therefore largely mitigated in MC inversions 
\cite[]{tarantola1982inverse}. %

We start the introduction of the SimRC by defining the covariance matrices between the parameters $\mathbf{m}$ following the prior distribution $\rho$, and the forward model responses $g(\mathbf{m})$: 

\begin{equation}
\mathrm{Cov}(g(\mathbf{m}),\mathbf{m})=\mathbf{E}[(\mathbf{m}-\mathbf{E}[\mathbf{m}])(g(\mathbf{m})-\mathbf{E}[g(\mathbf{m})])^T].
\label{eq:CovNonMatrix}
\end{equation}

\begin{figure}[H]
  \centering
\resizebox{15cm}{7.5cm}{%
\begin{tikzpicture}[sibling distance=19em, level distance=6em, 
  every node/.style = {shape=rectangle, rounded corners,
    draw, align=center,
    top color=white, bottom color=white}]]
    
  \node {$\mathrm{Cov}(g(\mathbf{m}),\mathbf{m})=\mathbf{E}[(\mathbf{m}-\mathbf{E}[\mathbf{m}])(g(\mathbf{m})-\mathbf{E}[g(\mathbf{m})])^T]$}
    child { node {Regression coefficient (RC) \\[1.5mm]$b_{j,i}:=\left(\mathrm{Cov}(g(\mathbf{m}),\mathbf{m}) \cdot \mathrm{Cov}(\mathbf{m},\mathbf{m})^{-1}\right)_{j,i}$} 
      child{
    	child{node {Standardized regression coefficient (SRC) \\[1.5mm]$\beta_{j,i}:=\frac{b_{j,i}}{\sigma(g_j(\mathbf{m}))}\cdot \sigma(m_i)$}}}}
    child { node {\baselineskip=2em Simplified regression coefficient (SimRC)\\[1.5mm]$\frac{\mathrm{Cov}(g_j(\mathbf{m}),m_i)}{\mathrm{Var(m_i)}}$}
      child { node {Prior-normed \\simplified  regression coefficient \\[1.5mm]$\frac{\mathrm{Cov}(g_j(\mathbf{m}),m_i)}{\sigma(m_i)}$}  
      	child { node {Correlation coefficient (CC) \\[1.5mm]$\frac{\mathrm{Cov}(g_j(\mathbf{m}),m_i)}{\sigma(m_i)\sigma(g_{j}(\mathbf{m}))}$} } } };
      	
   \node at (-9,0) {Units: $\left[g\right]\cdot\left[m\right]$}
   	child{ node {$\frac{\left[g\right]}{\left[m\right]}$}
   			child{ node {$\left[g\right]$}
   				child{ node {$\left[1\right]$}}}};
\end{tikzpicture}%
}
  \caption{Sensitivity measure candidates derived from the full prior distribution $\rho(\mathbf{m})$ and forward model $g$; $\mathbf{E}$ is the expectation operator with respect to $\rho(\mathbf{m})$; RC and SRC as used by \cite{saltelli2004sensitivity}.}
  \label{fig:flowchart}
\end{figure}
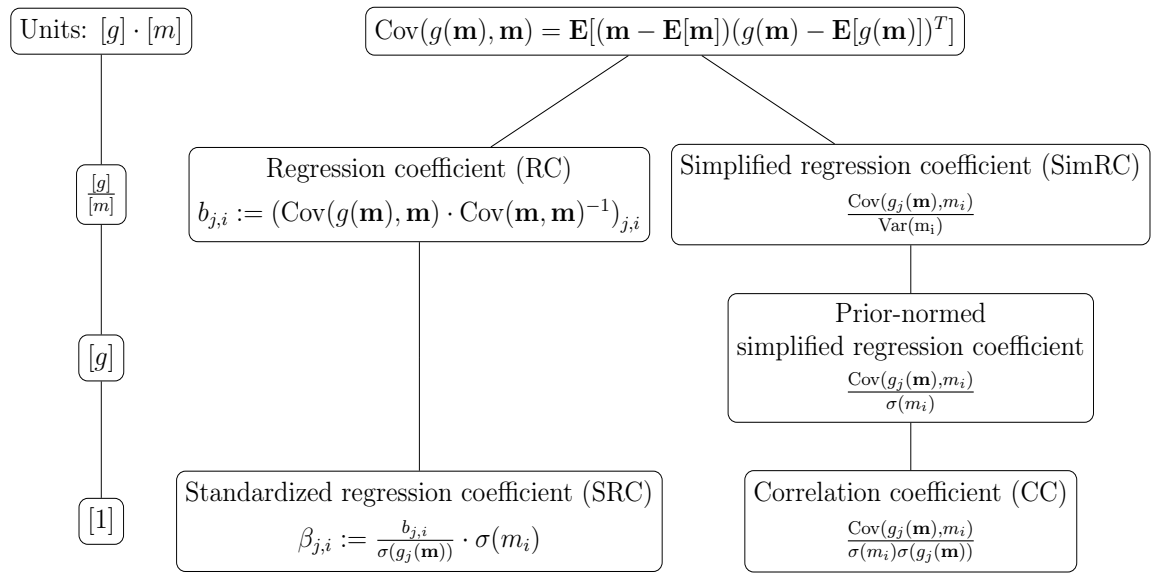

The covariances $\mathrm{Cov}(\mathbf{m},\mathbf{m})$ and $\mathrm{Cov}(g(\mathbf{m}),g(\mathbf{m}))$ are defined analogously. %
In Figure \ref{fig:flowchart}, we summarise the sensitivity measure candidates of this work that are derived from these covariances. %

From now on, we express probability distributions by an ensemble of Monte Carlo samples. %
Consequently, the covariances will be approximated as sample covariances. %
In terms of the ensemble matrices $\mathbf{M}$ and $\mathbf{G}$ 
the approximation of the covariance $\mathrm{Cov}(g(\mathbf{m}),\mathbf{m})$ reads

\begin{equation}
\mathbf{C}_{mg} := \mathbf{M'G'}^{T} \frac{1}{N-1} \approx \mathrm{Cov}(g(\mathbf{m}),\mathbf{m}), 
\label{eq:CovMatrixGeneral}
\end{equation}

where the primed matrices denote that for each parameter the 
parameter mean was subtracted from each parameter sample. %
Analogously, we define the (co-)variance matrices

\begin{equation}
\mathbf{V}_m := \mathbf{M'M'}^{T} \frac{1}{N-1} \approx \mathrm{Cov}(\mathbf{m},\mathbf{m}), 
\label{eq:Vm}
\end{equation}

and 

\begin{equation}
\mathbf{V}_g := \mathbf{G'G'}^{T} \frac{1}{N-1}  \approx \mathrm{Cov}(g(\mathbf{m}),g(\mathbf{m})). 
\end{equation}

In this notation, the sample approximations for the various coefficients given in Figure \ref{fig:flowchart} can be summarized as follows:

\begin{equation}
\mathrm{RC}_{j,i}=[\mathbf{V}^{-1}_m \mathbf{C}_{mg}]_{i,j} \qquad \mathrm{(regression \,\,coefficient)},
\label{eq:DefRC}
\end{equation}

and

\begin{equation}
\mathrm{SRC}_{j,i}=[\mathbf{V}^{-1}_m \mathbf{C}_{mg}]_{i,j} (\sigma_{m_{i}}/\sigma_{g_{j}})\qquad \mathrm{(standardized \,\,RC)}
\label{eq:DefSRC}
\end{equation}

by \cite{saltelli2004sensitivity}, where the index $j$ is used for responses and the index $i$ is used for model parameters. %
Consequently, the pair $ji$ could be read as sensitivity of response $j$ on parameter $i$. %

Additionally, we define

\begin{equation}
\mathrm{SimRC}_{j,i}=[\mathbf{C}_{mg}]_{i,j}/ \sigma^2_{m_{i}}\qquad \mathrm{(simplified \,\,RC)},
\label{eq:DefSimRC}
\end{equation}

and

\begin{equation}
\mathrm{CC}_{j,i}=[\mathbf{C}_{mg}]_{i,j}/ (\sigma_{m_{i}} \sigma_{g_{j}})\qquad \mathrm{(correlation \,\,coefficient)},
\label{eq:DefCC}
\end{equation}

where $\sigma^2_{m_{i}} := [\mathbf{V}_{m}]_{ii}$ and $\sigma^2_{g_{j}} := [\mathbf{V}_{g}]_{jj}$ are the sample variances. %

The comparison of various correlations, for example CC, can be easier than the 
comparison of covariances, as correlation is independent of 
units by normalization \cite[]{mosegaard1995monte}. %

\subsection{Discussion of the SimRC}

\paragraph{SimRC vs. differential sensitivity} 
Sensitivities characterise the influence of a parameter model 
on simulated measurement responses. %
Sensitivities are used for communicating 
geophysical inversion results as they support interpretation and 
possibly hint at adjustments of measurement design. %
Mostly, sensitivities are derived from a differential 
analysis, i.e. applying the difference quotient. %
For MC analysis, where Jacobian approximations are not readily available, 
we propose the usage of the SimRC as defined in the previous section. %
To illustrate similarities between SimRC and differential sensitivity, we now consider a linear forward model. %

\paragraph{Linear forward models}

We give the linear forward model of finite dimensions 
$g: \mathbb{R}^{n_{par}} \rightarrow  \mathbb{R}^{n_{obs}}$ the following general form: %

\begin{equation}
g(\mathbf{m})=\mathbf{A} \cdot \mathbf{m} + \mathbf{b},
\label{eq:line}
\end{equation}

with vector $\mathbf{b}$ for the intercept and  matrix $\mathbf{A} \in \mathbb{R}^{(n_{obs} \times n_{par})}$ for 
the slope. %
Arbitrary samples of $\mathbf{m}$ can be used to generate a 
$(\mathbf{m},g(\mathbf{m}))$-scatter plot, 
in which all points lie on the plane defined by Equation \ref{eq:line}.

Using the property of the sample covariance matrix that it is a quadratic form 
(\cite{horn2012matrix}; chapter 4.5.3), we get 

\begin{equation}
\mathrm{Cov}(\mathbf{m},g(\mathbf{m}))=\mathrm{Cov}(\mathbf{m},\mathbf{m}) \mathbf{A}^T.
\end{equation}

Here, $\mathrm{Cov}(\mathbf{m},\mathbf{m}) \in \mathbb{R}^{(n_{par} \times n_{par})}$ 
is the prior covariance matrix of the parameter vector. %
Inserting the Monte Carlo samples, we get the expression 

\begin{equation}
\mathbf{C}_{mg}=\mathbf{V}_m \mathbf{A}^T.
\label{eq:equation7}
\end{equation}

We now use (1) that quadratic covariance matrices are invertible (in the definition of RC), and 
(2) the general identity $\mathrm{Cov}(\mathbf{x},\mathbf{y})^{T}=
\mathrm{Cov}(\mathbf{y},\mathbf{x})$. %

Inserting Equation \ref{eq:equation7} in the definition of RC and SRC (Eq. \ref{eq:DefRC} and \ref{eq:DefSRC}), one derives

\begin{equation}
\mathrm{RC}_{j,i}=[\mathbf{A}]_{j,i}.
\label{eq:RC13} 
\end{equation}

In this linear case, this slope $\mathbf{A}$ is equal to both the 
differential sensitivity and the regression coefficient as used 
by \cite{saltelli2004sensitivity}. %

If the prior covariance matrix is a simple diagonal 
matrix, or equivalently no prior correlation is implemented, 
the relation $[\mathbf{C}_{mg}]_{i,j}=\sigma^2_{m_{i}} [\mathbf{A}]_{j,i}$ holds, and then can be inserted in the SimRC definition (Eq. \ref{eq:DefSimRC}), yielding the same expression as for RC$_{j,i}$ in Equation \ref{eq:RC13}. %

\paragraph{Non-diagonality} For non-diagonal prior covariance matrices, the SimRC 
differs from the general regression coefficient/differential 
sensitivity due to the simplified matrix inversion. %

In this work, we propose the SimRC for construction of sensitivity 
measures for two reasons. %
First, prior correlation has a strong effect 
on Bayesian inference and thus, it may be beneficial to incorporate it in a sensitivity measure. 
Monte Carlo samples of the prior distribution account for prior correlation. %
Likewise, the SimRC functions implicitly include prior correlation. %
Second, the computation of the SimRC is less expensive compared to 
the computation of the regression coefficient as given in 
Equation \ref{eq:DefRC}. %
For some cases, specifically for large models, the sample model variance matrix $\mathbf{V}_m=\mathbf{M'M'}^{T}/(N-1)$ (see Eq. \ref{eq:Vm}) may be singular. %
This singularity causes the non-existence of the matrix' inverse, disallowing the computation of RC and SRC. %

\paragraph{Non-linearity} For non-linear functions $g$, the slope at point $\mathbf{m}$ is computed 
using the difference quotient: 

\begin{equation}
\frac{\partial g_j (\mathbf{m})}{\partial m_i} \approx \frac{g_j (\mathbf{m}+\Delta m \mathbf{e}_i)-g_j (\mathbf{m})}{\Delta m},
\label{eq:jacob}
\end{equation}

for which the $i$th parameter was perturbed by $\Delta m$, and $\mathbf{e}_i$ 
is its unit vector. %
From the linear case, we can draw some conclusions regarding the 
relation of difference quotient and SimRC for general functions. %
The difference quotient and the SimRC should be approximately equal 
for prior parameters ranges that are in the order of $\Delta m_i$. %
Below, we will investigate the similarity of the SimRC and the 
difference quotient for two synthetic non-linear forward models. %
In practice the search space defined 
through the prior distribution is large compared to an infinitesimally 
small $\Delta m_i$. %
Thus, in the context of general non-linear forward models, the 
SimRC needs to be interpreted by comparing the prior 
parameter range to the non-linearity of the forward model. %

\subsection{Correlation and depth of investigation}

Often, an estimate for the depth of investigation (DOI) and sensitivity 
functions are provided along with inversion results. %
In probabilistic inversions, inversion results should be 
as least as reliable as the prior knowledge, regardless of how sensitive or insensitive the surface measurements are to the parameters. %
However, also for probabilistic inversions, sensitivity and DOI 
allow valuable insight into the nature of the inversion process. %
In particular, they can help to distinguish between parameters updated by Bayesian inference and parameters left more or less at their prior values. %
In this way, sensitivities and DOI may point out that an adjustment 
of measurement design is necessary. %

The SimRC gives a measure for the sensitivity of a measurement setup to 
a particular subsurface model. %
Thus, all available approaches for determining the DOI from sensitivity 
can be applied to the SimRC. %
However, for determining the DOI, a dimensionless sensitivity can have 
advantages. %
By normalizing the SimRC, one obtains the correlation between prior 
ensembles and forward response ensembles (Eq. \ref{eq:DefCC}). %
This correlation can be viewed as standardised SimRC corresponding 
to the standardised regression coefficient as used by \cite{saltelli2004sensitivity}. %

Due to its normalization, the correlation function has some beneficial 
properties when compared to the SimRC. %
First, the normalization makes sensitivities of different measurement signals 
comparable and easier to threshold. %
For example, the influence of signals from different electrode configurations 
in a direct current resistivity survey can be compared in 
terms of correlation. %
Second, for parameters with zero correlation to the measurement setup, no 
inversion update will be seen. %
Thus, zero correlation gives an intuitive measure for the DOI of 
Bayesian inference: the depth below which almost no correlation 
is found. %
One disadvantage of using correlation as sensitivity measure is 
that strong correlation can happen while there is small sensitivity (for a wide range of applicable parameter values). %
Thus, the information of how wide the range of equally applicable parameter values is, is lost by normalization. %

To complete the discussion of using the correlation for determining 
the DOI, we discuss the selection process. %
Due to two reasons a zero correlation DOI cannot be 
determined: (1) as for other sensitivity measures, the diffusive nature 
of energy propagation for the discussed measurement techniques prevents 
the correlation from reaching the zero function, and (2) the presence 
of undersampling in the MC ensemble may introduce spurious correlation. %

Due to reason (1) there is always some degree of arbitrariness 
in the DOI threshold selection. %
Thanks to its normalization and the resulting comparability between 
parameters, the correlation allows the definition of global DOI 
threshold values (for global DOI threshold values derived from 
differential sensitivity see \cite{oldenburg1999estimating} and 
\cite{vest2012global}). %

Regarding reason (2), as always in ensemble calculations, the 
ensemble has to be chosen large enough such that the spurious 
correlations are kept in the background noise. %

A useful approach accounting for fluctuations when determining the 
DOI from correlation sensitivity is the upwards cumulative sensitivity approach by \cite{vest2012global}. %
This approach can easily be applied to the CC functions 
(Eq. \ref{eq:DefCC}). %
By computing a cumulative sum, the possibly present spurious correlation 
is smoothed. %
To enable global thresholding, we introduce a normalization: %

\begin{equation}
\mathrm{CC}^{cum}_j(\mathbf{M,G})= \frac{\sum\limits_{i=j}^{n_{par}} |\mathrm{CC}_i|}{|\mathrm{CC}_{max}|},
\label{eq:CumCorr}
\end{equation}

with indices $j=1,...,n_{par}$, the correlation sensitivity 
$\mathrm{CC}_i$ for the $i$th layer, and the maximum correlation sensitivity 
$\mathrm{CC}_{max}$ for the vertical sequence of parameters. %

The upwards cumulative sensitivity approach will be analysed in the 
synthetic examples below. %

\section{Synthetic examples}

We now investigate the differences between differential sensitivity 
on the one hand, and 
the simplified regression coefficient (SimRC, Equation 
\ref{eq:DefSimRC}) and the correlation coefficient (CC, Equation \ref{eq:DefCC}) on the other hand. %
To this end, we look at the depth distribution of these parameters 
and use them to estimate a depth of investigation (DOI) for 
interesting cases. %
As outlined above, we expect that the differences between differential 
sensitivity and the SimRC are essentially determined by three 
parameters: (1) non-linearity of the physical model equations, (2) 
prior distribution correlations, and (3) the sampling bias of the ensemble. %
First, a linear toy model is used to illustrate the influence of 
sampling bias and prior correlation for the linear case. %
Second, for a non-linear toy model all three influencing parameters 
are investigated. %
Finally, we analyse the SimRC and correlation for a 
non-linear geophysical model and compare them to differential 
sensitivities. %

\subsection{Linear toy model}

We start with the relation of differential sensitivity and the SimRC 
using the simple case of a linear toy model. %
This toy model describes a surface measurement $S$ that is 
sensitive to a subsurface property $p$. %
The toy model is characterised by two key features: (1) a linear relation 
between the model parameters and the model response, and (2) an 
exponential decrease of the model response with depth. %
This exponential behaviour introduces a sensitivity that converges to 
zero with increasing depth but, similarly to diffusive geophysical 
models, a depth of truly zero sensitivity does not exist. %

We choose the explicit relation between measurement response $S$ and 
subsurface parameter $p$ as follows

\begin{equation}
S(z)=S_0 \cdot \frac{p}{p_0} \cdot \mathrm{e}^{-z/z_0},
\label{eq:responseLIN}
\end{equation}

with depth $z$, and constants $S_0$, $p_0$ and $z_0$. %
$S_0$ is the response of a subsurface layer with property $p_0$ at 
negligible depth. %
$z_0$ is the depth at which the response of a subsurface layer is reduced by a factor $1/\mathrm{e}$ compared to negligible depth. %
In general, these constants depend on the units of the corresponding 
physical variable. %
For simplicity, we set them to one. %

The total toy model response is given by an integration over the whole 
subsurface:

\begin{equation}
S^{total}=\int_0^{\infty} S(z) \mathrm{d}z.
\label{eq:StotalLin}
\end{equation}

We discretise Equation \ref{eq:StotalLin} using equally thick, discrete 
layers of constant $p$. %
The bottom layer of the model is chosen to extend to infinite depth. %
In the following, we investigate the two influences 
on the relation of the SimRC and differential sensitivity expected 
from theory: 
(1) different prior distribution probability density functions (PDFs), and (2) 
the influence of the ensemble size on the RC. %

We define some standard parameters that remain valid for all 
comparisons unless explicitly stated otherwise. %
First, each prior distribution is sampled drawing 100,000 samples. %
The standard deviations (STDs) of the respective forward responses 
are computed from the ensemble. %
We define a prior distribution with identical Gaussian prior PDFs for 
each discrete model layer: a mean of $p=3$ and a STD of 0.5. %
We compute the SimRC and correlation between the prior ensemble 
and the forward response ensemble computed by the linear toy 
model. %
We compare the SimRC function to the differential sensitivity 
for a homogeneous parameter model at $p=3$. %

\subsubsection{Gaussian prior PDF}

We study the SimRC and the correlation for the linear toy model 
using five different prior STDs: 0.01, 0.1, 0.5 , 1.0, and 1.5. %

\begin{figure}[H]
\centering
\includegraphics[scale=0.78]{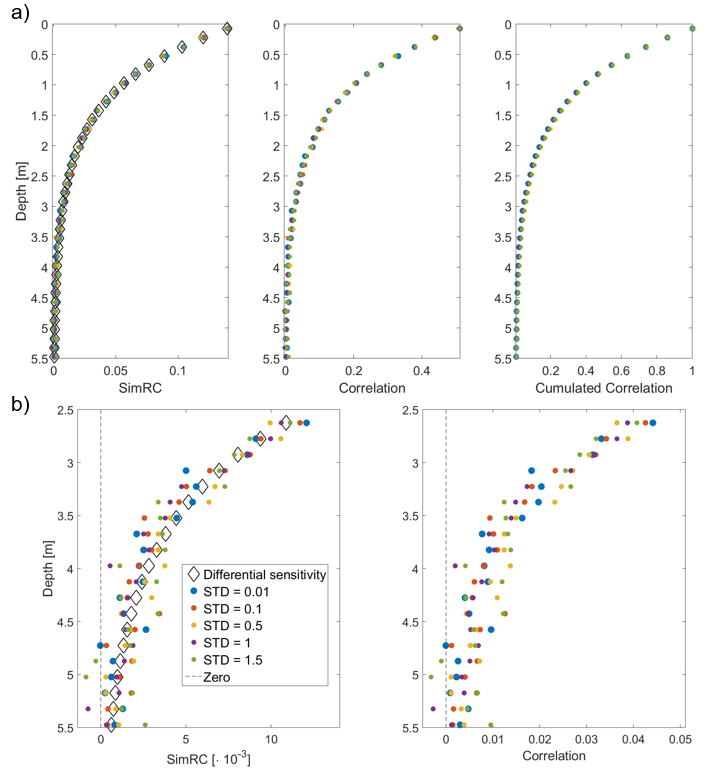}
\caption{a) Differential sensitivity, simplified regression coefficient (SimRC; left) and correlation (center, right) for five prior standard deviations (STD). Forward responses are computed using the linear toy model; b) Zoom-in of a).}
\label{fig:compareSTDintervalswithSens_Lin}
\end{figure}

Results for the SimRC, correlation, and differential sensitivity 
are shown in Figure \ref{fig:compareSTDintervalswithSens_Lin}a. %
As described by the toy model (Eq. \ref{eq:responseLIN}), the 
differential sensitivity asymptotically approaches the zero function 
with depth. %
As expected for linear model equations, all SimRC functions 
(Eq. \ref{eq:DefSimRC}) are very close to the differential sensitivity function. %
The shape of all correlation functions (Fig. 
\ref{fig:compareSTDintervalswithSens_Lin}a center) is similar 
to the differential sensitivity function. %
As a consequence, the cumulative correlation functions are very 
similar for all five prior distribution STDs. %

We now zoom in the results (Fig. 
\ref{fig:compareSTDintervalswithSens_Lin}b) in order to visualise 
sampling fluctuations. %
On average, the SimRC follows the differential sensitivity function. %
However, fluctuations around the differential sensitivity are 
present for all derived SimRC functions. %
Such fluctuations can be explained by the sampling bias. %

\begin{figure}[H]
\centering
\includegraphics[scale=0.38]{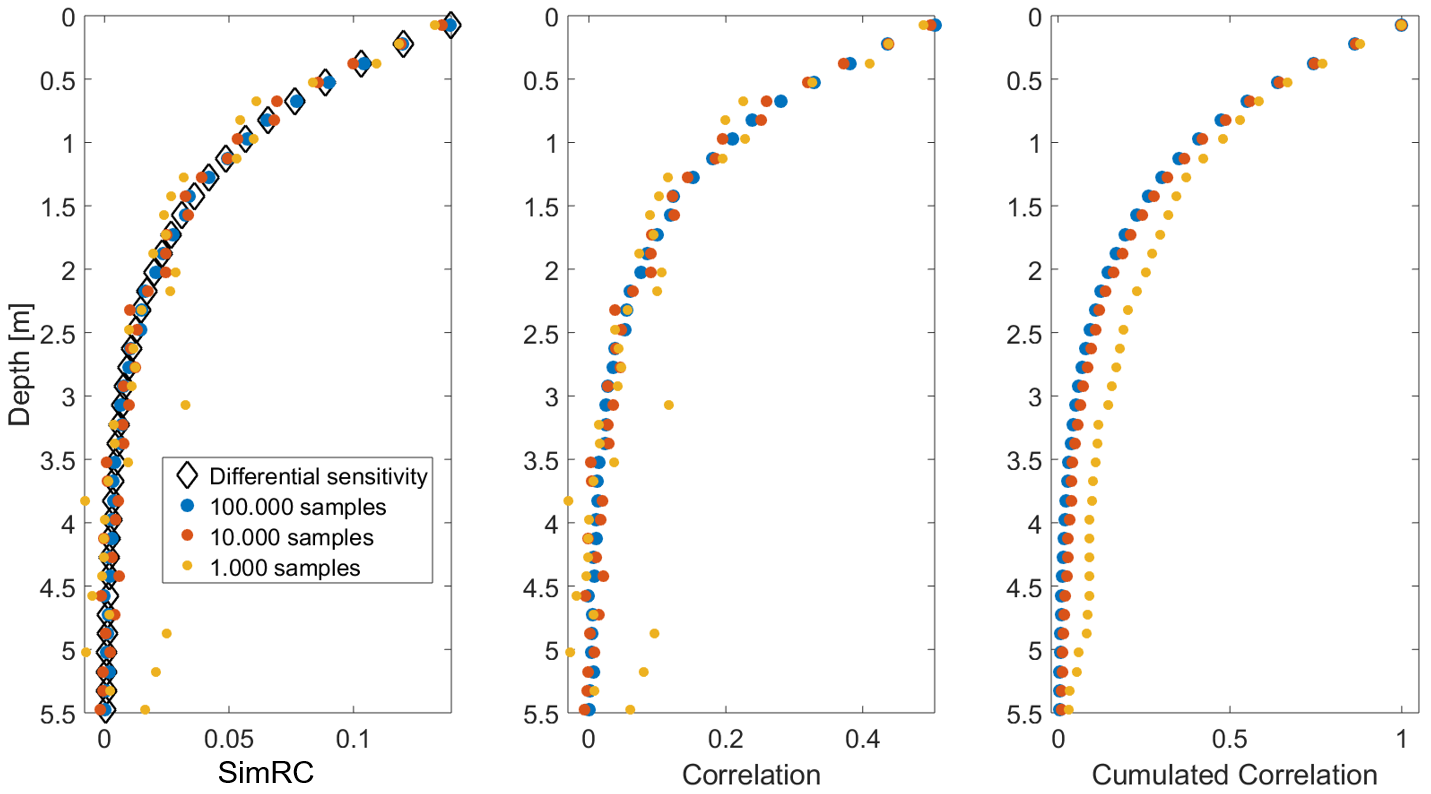}
\caption{Differential sensitivity, simplified regression coefficient (SimRC; left) and correlation (center, right) for three different ensemble sizes. Forward responses are computed using the linear toy model.}
\label{fig:smaplesize_Lin}
\end{figure}

To investigate the influence of the ensemble size on the 
SimRC and correlation functions, we re-sample the Gaussian prior 
PDF with a STD of 0.5 using the smaller ensemble 
sizes 10,000 and 1,000. %
Results are shown in Figure \ref{fig:smaplesize_Lin}. %
The comparison with differential sensitivity indicates the expected 
increase in the sampling error with decreasing ensemble size. %
The sampling error is also present in the different cumulative 
correlation functions. %
By normalizing and summing up, the fluctuations are smoothed and they 
stay visible as an offset generated by the larger absolute 
correlation values at depth. %

For an ensemble of only 1,000 samples, the undersampling effects 
become so pronounced that they prevent the interpretation of the 
SimRC as sensitivity. %
Thus, for this toy model, an ensemble size of at least 
10,000 is needed to use the SimRC as sensitivity, and subsequently 
for defining a DOI. %
However, it is important to note here that the minimal necessary number of samples in other studies may differ and always depend on the specific problem addressed. %

Theoretically, differential sensitivities can only take 
positive values for this toy model. %
Thus, negative sensitivity values are a result of sampling error. %
Generally, negative sensitivity values should only occur in regions 
below the DOI. %
For such profile sections to which measurements have low sensitivity, the correlation must be 
interpreted as zero, or, in the case of interpreting results from 
diffusive methods, as negligibly small. %

The DOI derived from the differential 
sensitivity using a threshold of 5 $\%$ of the maximum sensitivity 
is 3 meter. %
Since the SimRC values are almost equal to the differential 
sensitivities, they yield a DOI of 3 meter as well. %
For the correlations and cumulative correlation, we choose the 
thresholds that yield the same DOI of 3 meter in this simple 
linear case. %
For the correlations, this threshold correlation is 0.03. %
For cumulative correlation, this threshold is 0.05. %
For the more complicated examples, we will analyse the differences 
in DOI for these same thresholds. %

\subsubsection{Multivariate Gaussian prior PDF with vertical constraints}

To study the effect of prior correlation on the SimRC, we investigate 
four prior distributions with different correlation lengths. %
In general, prior correlation is implemented if a priori correlation between model parameters is assumed \cite[]{tarantola1982inverse}. %
We extend the separate Gaussian PDFs per parameter to a 
multivariate Gaussian PDF (e.g., \cite{bibby1979multivariate}). %
In this multivariate Gaussian PDF, parameters can be correlated via a 
prior covariance matrix \cite[]{hansen2006linear}. %
Here, we introduce a simple Gaussian shaped correlation, which is fully 
defined by a standard deviation of the prior parameters and a 
spatial correlation length (\cite{Gaspari1999}; Section 4.3).
As before, for all parameters the Gaussian mean is set to $p=3$ 
and a STD of 0.5 is chosen, creating a multivariate Gaussian 
prior distribution. %
We compute multiple ensembles for different prior distributions with increasing correlation length: 0 m, 0.2 m, 0.6 m and 1m. %
For the correlation length of zero meters, parameters 
are sampled independently, and thus, this prior distribution is equivalent 
to the uncorrelated Gaussian PDFs of the last section. %

\begin{figure}[H]
\centering
\includegraphics[scale=0.34]{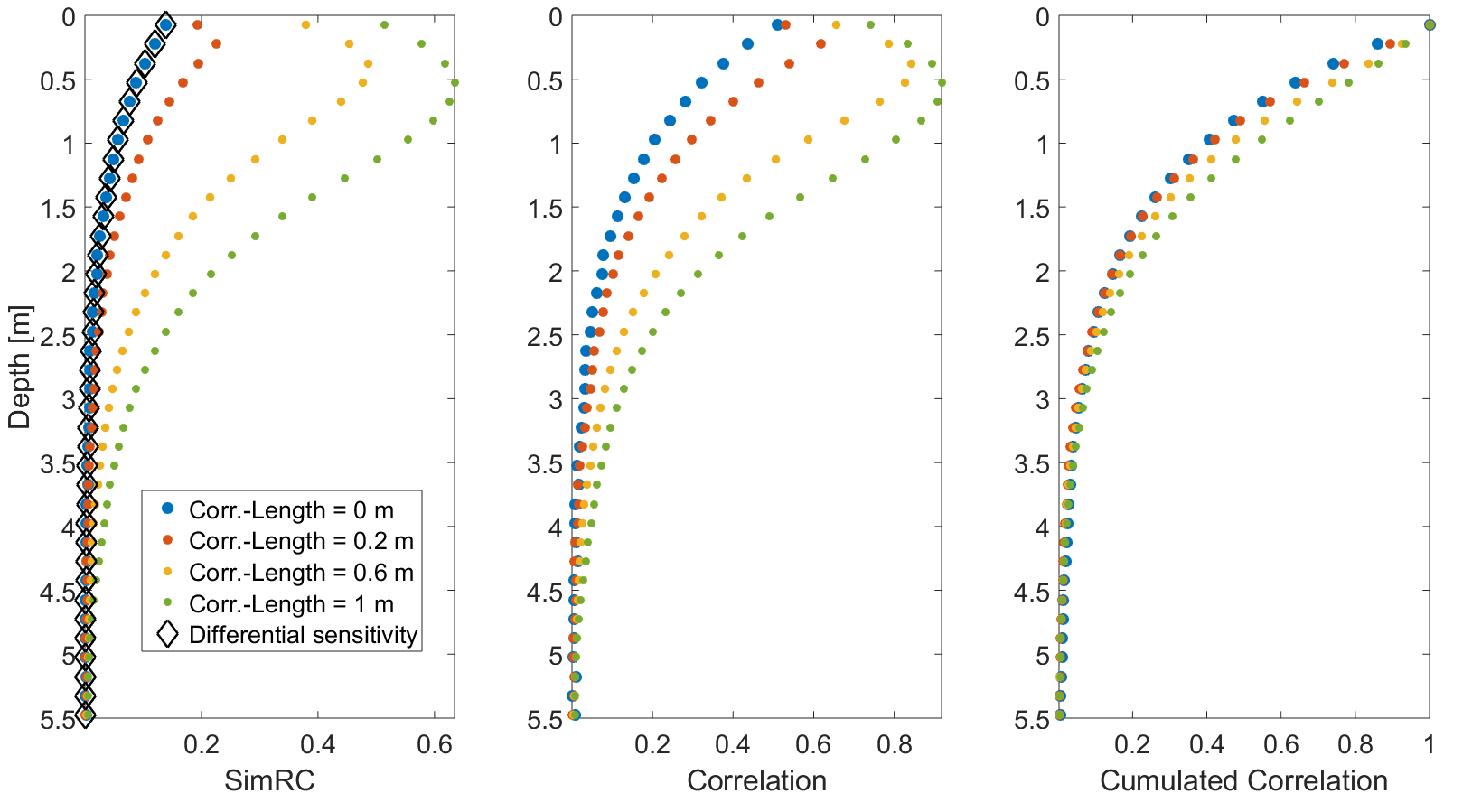}
\caption{Differential sensitivity, simplified regression coefficient (SimRC; left) and correlation (center, right) for four different correlation lengths of a uniform multivariate Gaussian prior distribution. Forward responses are computed using the linear toy model.}
\label{fig:Corrlength_Lin}
\end{figure}

SimRC results for different correlation lengths are shown in 
Figure \ref{fig:Corrlength_Lin}. %
The larger the prior correlation length, the larger the SimRC and 
consequently the correlation between prior ensemble and forward 
response ensemble. %
Prior correlation has the most obvious effect on the shallow parameters to which measurements are more sensitive. %
For the very shallowest parameters, correlation decreases again due 
to the decreasing number of neighbouring correlated layers. %

The difference between differential sensitivity and the SimRC 
caused by prior correlation has to be taken into account when interpreting SimRC functions as sensitivity. %
On the one hand, this is a deviation from the concept of 
differential sensitivity derived purely from partial 
derivatives. %
On the other hand, it may be beneficial, especially in Bayesian 
inversion methods, to incorporate the prior correlation into the sensitivity measure since it drives the Bayesian update. %
Of course, when interpreting the SimRC, it has to be kept in mind 
that the SimRC is not purely a property of the functional relationship
at hand. %
Instead, it is influenced by a mixture of the functional 
relationship and the prior correlation that is a human input. %
Finally, for the cumulative correlation functions, we can observe 
that the differences due to correlation are reduced by 
the normalization (Fig \ref{fig:Corrlength_Lin} right). %

Since an increase in correlation length leads to larger SimRC 
values, it is evident that it also enlarges a DOI estimate 
that is derived from SimRC and correlation analysis. %
As before, we use 5 $\%$ of the maximum sensitivity as the DOI 
threshold to compare DOIs estimated from differential sensitivity 
and SimRC functions. %
For the DOIs derived from correlation, we use a threshold correlation of 0.03. %
For the cumulative correlation we choose 0.05 as DOI threshold. %
These two threshold are set based on the results of the previous examples 
(see above). %

All DOI values are listed in Table 
\ref{tab:LinCorrlength}. %
The DOI for differential sensitivity is 3 meter. %
For the SimRC and the correlation, the derived DOIs are larger for 
larger correlation lengths. %
This behaviour has the same advantages and disadvantages as the form of the SimRC function as discussed before. %
The disadvantage is that the DOI is affected by input other than 
the function itself. %
On the other hand, the advantage is that the larger DOI properly 
reflects the sphere of influence of the Bayesian update. %
For the cumulative correlation, the DOIs are closer to the DOI of 
the differential sensitivity. %
We attribute this to the normalization that also brings the cumulative 
correlation curves closer together. %

\subsection{Non-linear toy model}

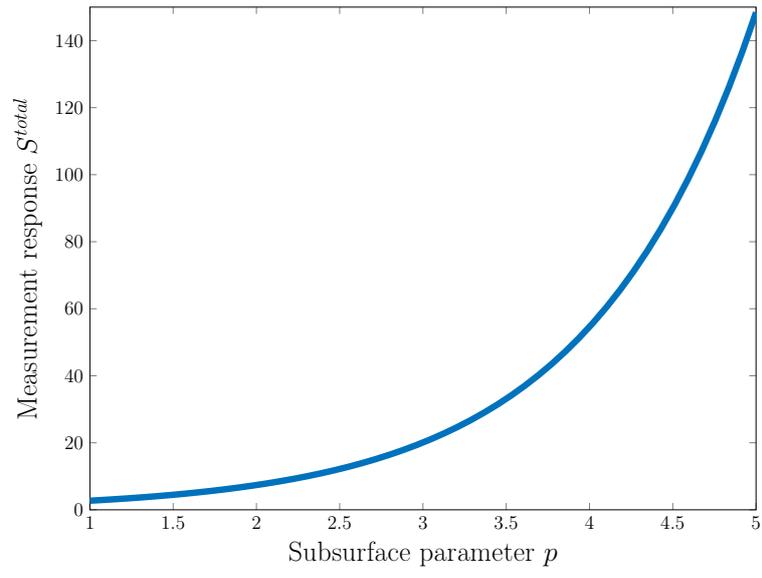
\begin{figure}[H]
\centering
\definecolor{mycolor1}{rgb}{0.00000,0.44700,0.74100}%
\resizebox{10cm}{7.5cm}{%
\begin{tikzpicture}

\begin{axis}[%
width=5.741in,
height=4.337in,
at={(1.206in,0.838in)},
scale only axis,
xmin=1,
xmax=5,
xlabel style={font=\color{white!20!black},font=\Large},
xlabel={Subsurface parameter $p$},
ymin=0,
ymax=150,
ylabel style={font=\color{white!20!black},font=\Large},
ylabel={$\text{Measurement response }  S^{total}$},
axis background/.style={fill=white}
]
\addplot [color=mycolor1, line width=4.0pt, forget plot]
  table[row sep=crcr]{%
1	2.71828182845905\\
1.08163265306122	2.94949111788976\\
1.16326530612245	3.20036640918951\\
1.24489795918367	3.47258043631491\\
1.3265306122449	3.76794821119589\\
1.40816326530612	4.08843912549383\\
1.48979591836735	4.43619008170061\\
1.57142857142857	4.81351974113148\\
1.6530612244898	5.2229439838115\\
1.73469387755102	5.6671926833358\\
1.81632653061224	6.1492279085515\\
1.89795918367347	6.67226367342275\\
1.97959183673469	7.23978736676288\\
2.06122448979592	7.85558300471835\\
2.14285714285714	8.5237564610426\\
2.22448979591837	9.24876284338497\\
2.30612244897959	10.0354361981285\\
2.38775510204082	10.8890217418364\\
2.46938775510204	11.8152108342135\\
2.55102040816327	12.8201789257675\\
2.63265306122449	13.9106267331906\\
2.71428571428571	15.0938249170008\\
2.79591836734694	16.3776625593359\\
2.87755102040816	17.770699765131\\
2.95918367346939	19.2822247374005\\
3.04081632653061	20.9223157071823\\
3.12244897959184	22.7019081310646\\
3.20408163265306	24.6328676043435\\
3.28571428571429	26.7280689759649\\
3.36734693877551	29.0014821927583\\
3.44897959183673	31.4682654453345\\
3.53061224489796	34.1448662367092\\
3.61224489795918	37.0491310475333\\
3.69387755102041	40.2004243291359\\
3.77551020408163	43.6197576177749\\
3.85714285714286	47.3299296309774\\
3.93877551020408	51.3556782800744\\
4.02040816326531	55.7238456124879\\
4.10204081632653	60.4635567835381\\
4.18367346938776	65.6064142510805\\
4.26530612244898	71.1867084877852\\
4.3469387755102	77.2416466160007\\
4.42857142857143	83.8116004896458\\
4.51020408163265	90.9403758772393\\
4.59183673469388	98.6755045408695\\
4.6734693877551	107.068561158566\\
4.75510204081633	116.175507203183\\
4.83673469387755	126.057064070643\\
4.91836734693878	136.779117945394\\
5	148.413159102577\\
};
\end{axis}
\end{tikzpicture}%
}
\caption{Exponential dependency of $S^{total}$ on the subsurface parameter $p$.}
\label{fig:toymodel}
\end{figure}

To study differences between differential sensitivity and the SimRC for 
a non-linear parameter dependence, we introduce a simple but strongly 
non-linear toy model. %
We adjust the linear toy model from the previous section 
by turning the linear relation between the model response and the 
subsurface parameter $p$ into an 
exponential one (Figure \ref{fig:toymodel}):

\begin{equation}
S(z)=S_0 \cdot \mathrm{e}^{p/p_0} \cdot \mathrm{e}^{-z/z_0}.
\label{eq:nonlinmodel}
\end{equation}

The total response is computed for a discretised model analogous 
to the linear case. %
As for the linear example, we compute SimRC and correlation of prior and 
forward response ensemble for different prior distributions and ensemble sizes. %
As before, the default prior distribution is a Gaussian 
PDF with mean $p=3$ and STD of 0.5, sampled using an ensemble size 
of 100,000, unless stated otherwise. %

\subsubsection{Gaussian prior PDF}

We study the effect of the non-linearity between the forward 
model response and the prior distribution by using five prior distributions 
differing in STD. %
The STDs are 0.01, 0.1, 0.5 , 1.0, and 1.5. %
For larger STD, the ensemble will be more non-linear. %

\begin{figure}[H]
\centering
\includegraphics[scale=0.7]{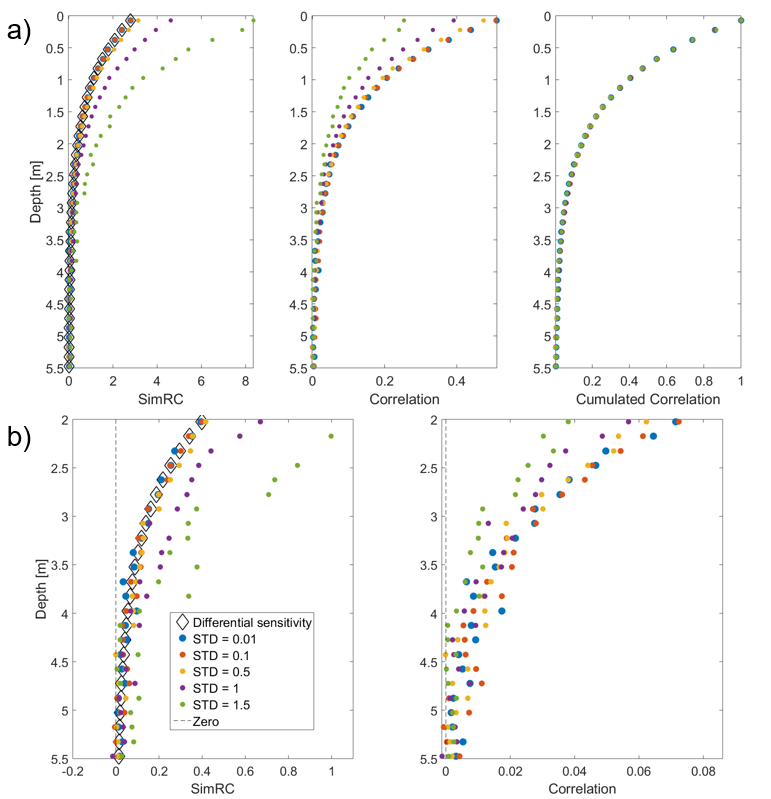}
\caption{a) Differential sensitivity, simplified regression coefficient (SimRC; left) and correlation (center, right) for five prior standard deviations (STD). Forward responses are computed using the non-linear toy model; b) Zoom-in of a).}
\label{fig:compareSTDintervalswithSens}
\end{figure}

Results are shown in Figure \ref{fig:compareSTDintervalswithSens}. %
For the two smallest STDs, the SimRC function 
(Eq. \ref{eq:DefSimRC}) is close to the differential 
sensitivities (Fig. \ref{fig:compareSTDintervalswithSens}a left). %
In the more shallow profile sections, we observe that the 
SimRC increases for the larger STDs. %
As the parameter-response relation is exponential, this 
behaviour is expected. %
For the larger STDs, the curvature of the non-linearity of the 
forward model leads to regression coefficients that are larger 
than the slope at the mean of the prior ensemble. %
If the curvature had a different sign, such that it would 
reduce the slope of the function, it would lead to SimRC values 
that are smaller than differential sensitivity. %

In the correlation functions (Fig. \ref{fig:compareSTDintervalswithSens}), 
we see that an increase in prior STD causes a 
decrease in the correlation for the more shallow model parameters. %
As correlation is a measure for the linearity of the relation 
between two 
random variables, this decrease of correlation is expected for the 
non-linear toy model. %
For correlation, any curvature would lead to a decrease. %
The cumulative correlation function is similar for all investigated 
prior STDs (Fig. \ref{fig:compareSTDintervalswithSens}a right). %
The integration over depth almost removes the effects of the non-linearities in 
the correlation functions. %
Thus, a DOI estimation derived from cumulative correlation functions would be almost identical for all STDs. %

For the SimRC, fluctuations around the differential 
sensitivity function are observed especially in the lower most 
parts of the profile (Fig. 
\ref{fig:compareSTDintervalswithSens}b). %
Fluctuations are more pronounced for larger STDs. %
Thus, a strong non-linearity in parameter-response relation 
requires a larger ensemble size to achieve a similarly low degree 
of ensemble bias as seen for less non-linear relations. %

Turning to the DOIs, we use the same threshold as for the 
linear case. %
For the differential sensitivity, this results in a DOI of 3 m. %
All DOI values are listed in Table \ref{tab:Nonlin_STD}. %
SimRC DOIs are close to the 3 meter DOI derived for the 
differential sensitivity function, regardless of the prior distribution STD. %
For the DOI derived from the correlation functions, a decrease in DOI is 
observed for an increasing prior STD. %
This effect can be traced back to the general decrease of 
correlation for an increasing non-linearity (see above). %
As for the linear example, the cumulative correlation is 
closer to the DOI derived for differential sensitivity by 
normalization. %

\subsubsection{Multivariate Gaussian prior PDF with vertical constraints}

As for the linear case, we now introduce prior correlation 
by turning the Gaussian priors from the previous section into 
multivariate Gaussian priors. %
The SimRC is analysed for these correlation lengths: 
0 m, 0.2 m, 0.6 m and 1m. %

\begin{figure}[H]
\centering
\includegraphics[scale=0.38]{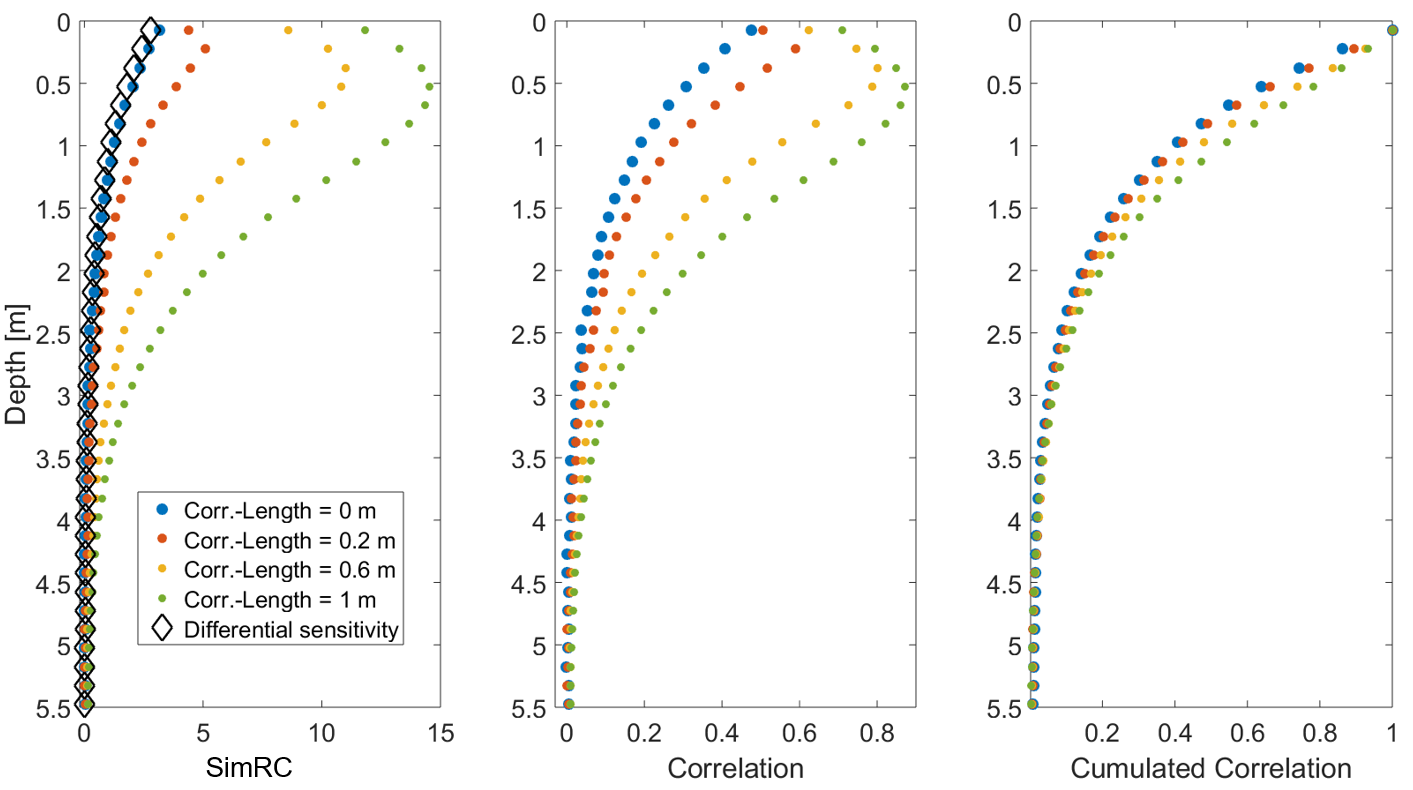}
\caption{Differential sensitivity, simplified regression coefficient (SimRC; left) and correlation (center, right) for four different correlation lengths of a uniform multivariate Gaussian prior distribution. Forward responses are computed using the non-linear toy model.}
\label{fig:Corrlength}
\end{figure}

Results are shown in Figure \ref{fig:Corrlength}. %
By comparing the SimRC to differential sensitivity, we see 
two influences on the SimRC functions: 
(1) a relatively small offset arising from the non-linearity in the model response, 
and (2) the relatively larger influence of the prior correlation. %
As for the linear case, the SimRC and correlation in the 
shallow profile parts increase with increasing correlation length 
as parameters here are most sensitive to the toy model. %

Since the sensitivity measures for this case are very similar case 
with correlation, we do not explicitly show the DOI values. %
As in the linear case, the DOI estimates for the SimRC in the region 
of enhanced correlation are larger compared to the 
DOI estimate from differential sensitivity. %

\subsubsection{Uniform prior PDF}

We study the applicability of SimRC and correlation analysis 
of uniform PDFs as a simple non-Gaussian case. %
We study the non-linearity of the toy model response by 
increasing the size of the prior interval. %
In particular, we derive the SimRC and correlation functions for five 
different uniform prior intervals around a mean of $p=3$: $\pm$0.05, 
$\pm$0.5, $\pm$2.5,$\pm$5.0, and $\pm$7.5. %

\begin{figure}[H]
\centering
\includegraphics[scale=0.38]{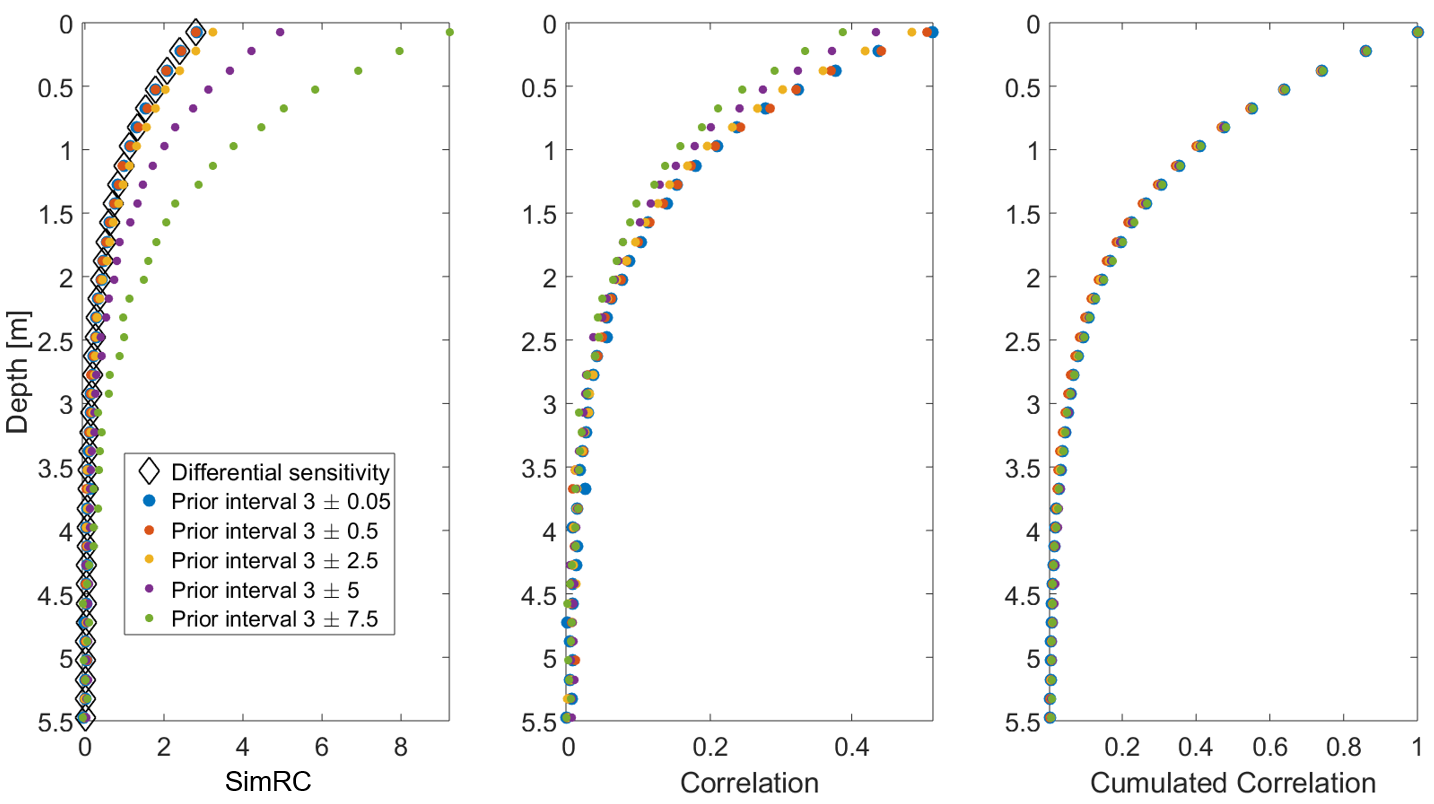}
\caption{Differential sensitivity, simplified regression coefficient (SimRC; left) and correlation (center, right) for five uniform prior intervals. Forward responses are computed using the non-linear toy model.}
\label{fig:uniformprior}
\end{figure}

Results are shown in Figure \ref{fig:uniformprior}. %
The SimRC and correlation functions are similar to the 
Gaussian case. %
Again, for the two smallest prior intervals, the differential sensitivity 
approximately equals the computed SimRC. %
As for the Gaussian case, correlation 
decreases with an increase of non-linearity on the investigated 
parameter-response interval. %
The cumulative correlation functions are closer together for 
all investigated prior distribution PDFs. %
This simple example shows that the usage of SimRC and correlation 
as sensitivity measures is not strictly restricted to Gaussian PDFs. %

\subsection{Frequency-domain electromagnetic model}

Frequency-domain electromagnetic (FDEM) induction data show diffusive 
energy propagation in the subsurface when low frequencies are used for 
surveying. %
Therefore, measurement sensitivity and DOI are quantities of interest when 
FDEM are processed and discussed. %

FDEM data are sensitive to the electrical conductivity (EC) of the 
subsurface. %
During FDEM surveys, an electromagnetic field is generated in a transmitter coil. %
The field propagates into the subsurface and generates eddy currents 
in conductive material. %
These eddy currents subsequently generate a secondary field 
that is recorded by one or multiple receiver coils. %
The magnitude of the secondary field is usually expressed in parts-per-million [ppm] with respect to the primary field. %
This magnitude can be related to EC using 
the non-linear formulas given by \cite{ward1988electromagnetic}. %

\begin{figure}[H]
\centering
\includegraphics[scale=0.35]{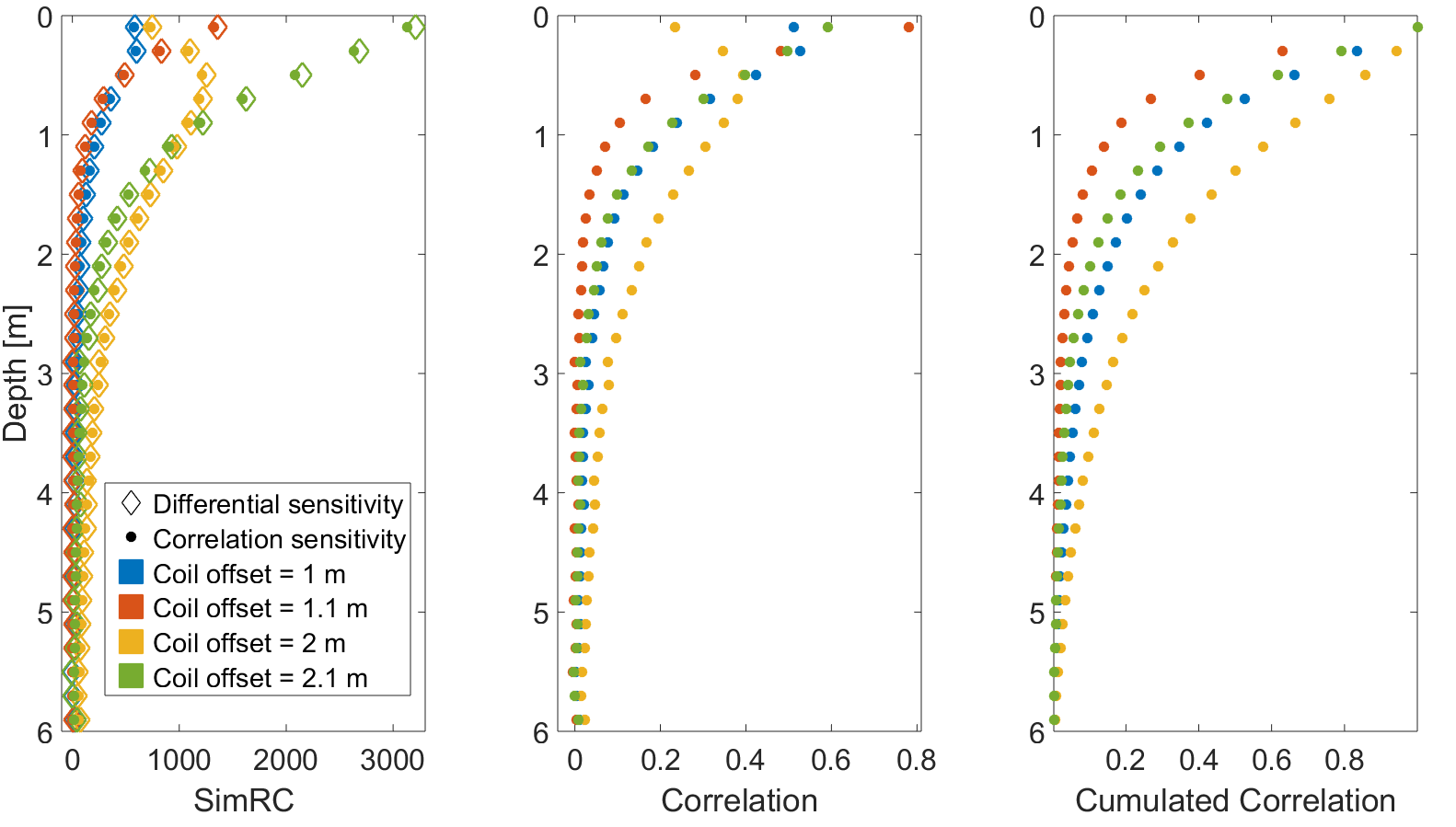}
\caption{Differential sensitivity, simplified regression coefficient (SimRC; left) and correlation (center, right) for a uniform Gaussian prior distribution and the different simulated measurement signals for four FDEM receiver coils with different distance to the transmitter coil.}
\label{fig:EMcompare}
\end{figure}

For one-dimensional forward modelling of FDEM data, we use the code provided 
by \cite{hanssens2019frequency}. %
We simulate data for one transmitter and two receiver coils, 
each of the receivers simulated at two different lateral offsets. %
The first receiver is simulated with a horizontal co-planar coil setup at two locations, in particular at 1 m and 2 m offset to the transmitter coil. %
The second receiver is simulated with a perpendicular coil setup at 1.1 m and 2.1 m offset from the transmitter coil. %
This way, simulated differential sensitivity functions 
(Fig. \ref{fig:EMcompare}) of the four signals differ (1) in magnitude 
due to the offset variations, and (2) in shape due to the difference in 
setup geometry. %

\begin{figure}[H]
\centering
\includegraphics[scale=0.33]{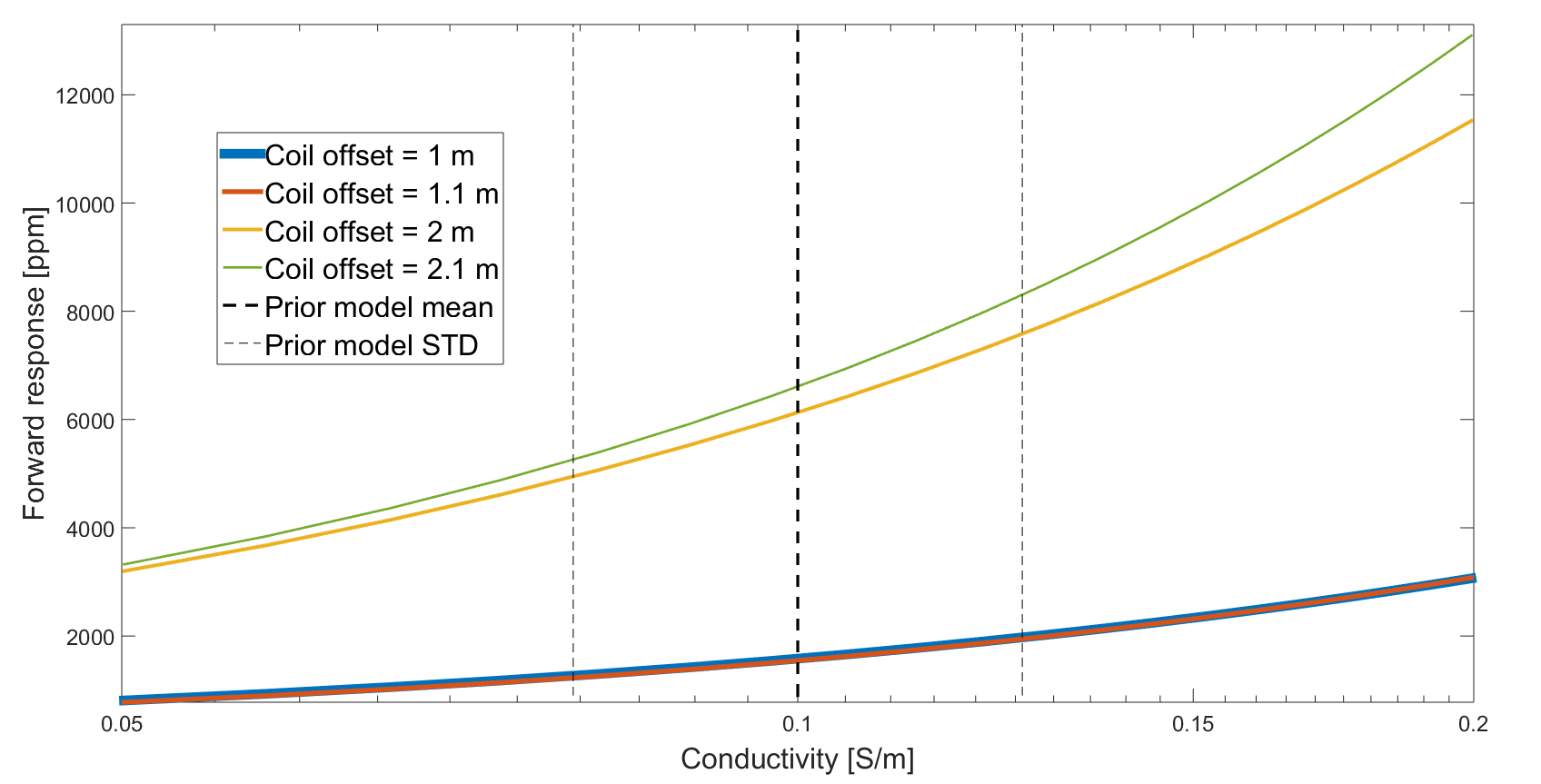}
\caption{Forward response for the FDEM forward model around the Gaussian 
prior distribution mean.}
\label{fig:EMlin}
\end{figure}

Data are simulated for a discretised subsurface for which each layer is 
20 cm thick. %
To prevent unphysical (negative) parameter values, prior 
model parameters are defined through log-normal PDFs for the subsurface parameters. %
The prior ensemble has a size of 100,000. %
Around the selected prior mean (EC of 100 mS/m), the forward response shows a 
slightly non-linear behaviour (Fig. \ref{fig:EMlin}). %
For the coils with larger offset, the non-linearity is more pronounced 
than for the two smaller offset measurements. %

The statistical parameters for SimRC analysis for all four 
measurement signals are shown in Figure \ref{fig:EMcompare}. %
The SimRC and the differential sensitivity 
are almost identical. 
After our analysis for the synthetic models, differences between the
two quantities can be associated to two potential sources: (1) the
non-linearity of the forward equation as shown in Figure \ref{fig:EMlin}, 
and (2) and undersampling bias in the ensemble. %
The most non-linear signal, the simulated response for the perpendicular 
2.1 m coil, accordingly shows the largest difference of differential sensitivity of the uniform parameter model with an EC of 100 mS/m and 
the SimRC. %

As different signals are compared here, this example gives a good 
illustration of the normalizing effect implicitly 
included in the correlation derived from the SimRC. %
Contrasting SimRC to correlation (Fig. \ref{fig:EMcompare} 
left and center), correlation functions of all four simulated 
measurements signals are directly comparable for the simulated EC 
profile. %

A comparison of the DOIs is based on the thresholds from the 
synthetic examples. %
DOI values are listed in Table \ref{tab:DOI_EMmodel}. %
DOIs for differential sensitivity and SimRC are about equal, as 
non-linearity is in the responses around the ensemble mean is 
comparable small. %

In general, DOIs derived from SimRC and correlation are in close 
agreement with the DOIs derived from differential sensitivity. %
The largest difference occurs for the horizontal co-planar coil at 
offset 2.0 m. %
For this configuration, the largest DOIs are seen. %
Looking at Figure \ref{fig:EMcompare}, we see that all sensitivity 
measures already close to zero in this range. %
Thus, a small change in the threshold would lead to considerably 
different DOIs. %

All in all, the SimRC and correlation functions yield sensitivity
measures and DOIs in agreement with differential sensitivity for 
this FDEM forward model. %

\section{Conclusion}
\label{sec:conclusion}

In this work, we introduce the simplified regression coefficient 
(SimRC). %
Additionally, we propose the SimRC and correlation of prior ensemble 
and forward response ensemble as sensitivity 
measures and for DOI estimation. %
Both measures are closely related to differential 
sensitivities of measurement variables to model parameters. %
For linear problems, we showed that SimRC and differential 
sensitivity are equal if no correlation is 
implemented in the prior distribution. %

For Monte Carlo inversion methods, the SimRC is computed from the 
readily available covariance matrix. %
Thus, no additional forward model runs are needed. %
This makes the SimRC a computationally attractive alternative to 
the Jacobians needed to compute differential sensitivity. %

We analyse the use of the SimRC analysis for three synthetic examples:
a linear, a simple non-linear, and a frequency-domain electromagnetic
forward model. %
For each forward model, we analyse the following influences on the
SimRC function: (1) forward model non-linearities, (2) a priori
defined (vertical) correlation between the model parameters, and (3)
ensemble fluctuations caused by undersampling. %
All three influences cause differences between differential
sensitivity and the SimRC. %

Regarding the non-linearity of the forward model, differences between
differential sensitivity and SimRC increase with an increase in the
non-linearity of our synthetic forward model. %
The deviation between differential sensitivity and the SimRC depends
on the curvature of the forward model. %
For correlation, the non-linearity leads to smaller absolute
correlations. %
The influence of non-linearity on DOI estimation was minor, the SimRC,
correlation, and cumulative correlation yield DOIs similar to the
differential sensitivities. %

While the influence of non-linearity is small-sized, the SimRC and
correlation may have an advantage over differential sensitivities for
some specific forward models. %
For our realistic exponential non-linearity, they account for the
average of non-linearity of a whole interval of the definition range
of the forward model. %
This averaging property can be even more vital for other forward
models. %
For example, for alternating forward models, local partial derivatives
may give a wrong impression of a function, while the ensemble used to
compute the SimRC and correlations has the ability to capture the
forward model behaviour across the whole prior range. %

Regarding a priori correlation, differences of differential
sensitivity and the SimRC are more pronounced than for
non-linearity. %
Two main deviations can be observed. %
First, the SimRC and correlation are much larger in the most sensitive
area of the subsurface. %
Second, the DOI becomes larger, as regions of the subsurface that were
previously not sensitive become sensitive through correlation. %

The a priori correlation introduces an important difference between
differential sensitivities and the SimRC. %
While differential sensitivity is fully determined by the underlying
forward model, the SimRC, by including effects of correlation between
prior distribution parameters, becomes dependent on a second human input. %
This is a disadvantage of the SimRC (and general stochastic Monte Carlo methods) and it always has to be kept in
mind when analysing its values. %
On the other hand, the dependency on prior correlation can be
desirable since this correlation is important for the Bayesian
update. %
Thus, when insight into the update process is the target of the
sensitivity study, the SimRC functions gives valuable
extra-information compared to differential sensitivity. %
If a separate analysis of pure measurement sensitivity is desired, one
could additionally compute the (computationally more expensive)
regression coefficient. %

Regarding ensemble fluctuations, as for all ensemble methods, the
overall size of the ensemble has to be large enough for the problem at
hand. %
Here, this means that small ensemble size leads to ensemble bias which
hinder an sensible assessment of the sensitivity and a clear
determination of the DOI. %
In these cases, it is likely that the ensemble is also too small for
the actual Bayesian update. %

Finally, applying the SimRC analysis to a geophysical forward model,
we obtain similar information from differential sensitivity and the
SimRC. %
After checking that non-linearities in the investigated
frequency-domain electromagnetic model are comparably small for the
investigated prior range and if no prior correlation is implemented,
the SimRC can be used and interpreted as a (classical) differential
sensitivity. %

Overall, we recommend using DOI estimations from the SimRC for
geophysical parameter estimations using Bayesian inference. %
For correlation functions, obtained by normalizing the SimRC, global
DOI thresholds can be introduced. %
When the influences of prior correlation and non-linearity are kept in
mind, the SimRC and correlation yield a computationally attractive
sensitivity adapted to judging the sphere of influence of a Bayesian
update. %

Analogous to differential sensitivity, the SimRC can be extended to 2-D and 3-D models in a straightforward manner. 
In the future, the SimRC should be applied and tested for such  models beyond the 1-D approximation, further analysing its advantages over Jacobian and RC computations. %
The computational advantage over the Jacobian becomes especially beneficial for highly-parametrised models resulting in expensive forward model runs. %
Additionally, for cases where the number of model parameters exceeds the number of Monte Carlo samples, the sample model variance matrix may be singular, causing the RC and SRC to be unavailable while the SimRC is still available. %

While in this work the SimRC was solely compared to differential sensitivity and thereof derived simple measures for the DOI, more sophisticated DOI estimation approaches as outlined in the introduction should be compared to the DOI estimations derived from SimRC and the correlation functions. %
In particular, the normalized Kullback-Leibler divergence and the correlation functions could be compared for multiple synthetic cases to further investigate three main questions: (1) how the Kullback-Leibler divergence relates to the SimRC functions, (2) if more generally valid DOI thresholding values can be defined, and (3) how a DOI estimate from Kullback-Leibler divergence relates to the SimRC DOI estimates derived for the measurement signals. %
For such a comparison, it must be acknowledged that SimRC does not formally provide an overall DOI. %
However, considering the SimRC function of the measurement signal showing the \textit{deepest} significant sensitivity to the subsurface parameters may be adequate for a meaningful comparison to the Kullback-Leibler divergence function. %

%

\section*{Acknowledgements}

This project has received funding from the European Union’s EU Framework Programme for Research and Innovation Horizon 2020 under Grant Agreement No 721185. %
          
%
\bibliography{DOIbib}
\bibliographystyle{apalike}

\newpage

\listoftables

\newpage

\begin{table}[H]
\centering
\caption{Depth of investigations for regression coefficient (SimRC; DOI threshold=5 $\%$ of maximum sensitivity), correlation (DOI threshold=0.01), and cumulative correlation (DOI threshold=5 $\%$) for four different correlation lengths for the prior distribution. Forward responses are computed using the linear toy model. The DOI of the differential sensitivity (DOI threshold=5 $\%$ of maximum sensitivity) is 3.0 m.}
\begin{tabular}{ |c|c|c|c| } 
 \hline
 Corr.-Length [m] & DOI SimRC [m] & DOI Corr. [m] & DOI Cum. Corr. [m] \\
 \hline 
 0.0 & 3.15 & 2.85 & 3.15\\  
 0.2 & 3.30 & 3.30 & 3.00\\
 0.6 & 3.60 & 3.75 & 3.15\\
 1.0 & 4.05 & 4.35 & 3.30\\
 \hline
\end{tabular}
\label{tab:LinCorrlength}
\end{table}

\begin{table}[H]
\centering
\caption{Depth of investigations for regression coefficient (SimRC; DOI threshold=5 $\%$ of maximum sensitivity), correlation (DOI threshold=0.01), and cumulative correlation (DOI threshold=5 $\%$) for five different standard deviations (STD) of the prior distribution. Forward responses are computed using the non-linear toy model. The DOI of the differential sensitivity (DOI threshold=5 $\%$ of maximum sensitivity) is 3.0 m.}
\begin{tabular}{ |c|c|c|c|c| } 
 \hline
 STD & DOI SimRC [m] & DOI Corr.[m] & DOI Cum. Corr. [m]\\
 \hline 
 0.01 & 3.15 & 2.85 & 3.00\\  
 0.1 & 3.15 & 2.85 & 3.15\\
 0.5 & 3.00 & 3.00 & 3.00\\
 1.0 & 3.00 & 2.55 & 3.15\\
 1.5 & 2.85 & 2.40 & 3.15\\
 \hline
\end{tabular}
\label{tab:Nonlin_STD}
\end{table}

\newpage

\begin{table}[H]
\centering
\caption{Depth of investigations for regression coefficient (SimRC) and differential sensitivity (both with DOI threshold=5 $\%$ of maximum sensitivity), correlation (DOI threshold=0.01), and cumulative correlation (DOI threshold=5 $\%$) for the four simulated FDEM forward responses corresponding to the coil offsets. The last row corresponds to the overall DOI (maximum of the DOIs for the separate signals).}
\begin{tabular}{|p{2.5cm}|p{2.5cm}|p{2.5cm}|p{2.5cm}|p{2.5cm}| } 
 \hline
 Coil Offset [m] & DOI Diff. Sens. [m]  & DOI SimRC [m] & DOI corr. [m] & DOI cum. corr. [m]\\
 \hline 
 1.0 & 3.4 & 3.6 & 3.2 & 4.0 \\  
 1.1 & 1.6 & 1.6 & 1.8 & 2.2\\
 2.0 & 6.0 & 5.6 & 4.8 & 4.8\\
 2.1 & 2.8 & 3.0 & 3.0 & 3.2 \\
 \hline
\end{tabular}
\label{tab:DOI_EMmodel}
\end{table}

\newpage
\listoffigures
 
\newpage

\newpage

\newpage

\newpage

\newpage

\newpage

\newpage

\newpage

\newpage

\newpage

\end{document}